\documentstyle[preprint,aps,epsf]{revtex}
\begin{document}
\draft

\title{The Quantum Josephson Hamiltonian In The Phase Representation} 
\author{J.R.~Anglin$^1$, P. Drummond$^2$ and A. Smerzi$^3$}
\address{
1)Institute for Theoretical Atomic and Molecular Physics, 
Harvard-Smithsonian Center for Astrophysics, 
60 Garden Street,
Cambridge, MA 02138
USA\\
2)Physics Department, University of Queensland,  St Lucia 4072, 
Queensland, Australia\\
3) Istituto Nazionale di Fisica della Materia and  
International School for Advanced Studies,
 via Beirut 2/4, I-34014, Trieste, Italy
}
\date{\today}
\maketitle
\begin{abstract}
The quantum Josephson Hamiltonian of two weakly linked 
Bose-Einstein condensates is written in an overcomplete phase representation, 
thus avoiding the problem of defining a Hermitian phase operator. 
We discuss the limit of validity of the standard, 
non-rigorous Mathieu equation, 
due to the onset of a higher order $\cos 2 \phi$ term in the Josephson 
potential, and also to the overcompleteness of the representation
(the phase $\phi$ being the relative phase between the two condensates).
We thereby unify the Boson Hubbard and Quantum Phase models.
\end{abstract}
\pacs{PACS: 74.50.+r,03.75.Fi}

Recent developments on the engineering of 
Bose-Einstein condensates (BEC's) \cite{anderson98,andrews97:misc}
are suggesting new scenarios for searching 
macroscopic quantum coherence phenomena. 
Most of the dynamical regimes investigated experimentally 
have been
understood within the Gross-Pitaevskii (GP) framework \cite{dalfovo99}. 
GP is essentially a ``classical'' theory, since it approximates the  
quantum bosonic field as a complex order parameter
\cite{pitaevskii61}: the new challenge is the systematic study of
regimes where quantum fluctuations cannot be ignored.

Quantum effects play an important role
in a ``mesoscopic'' Josephson junction \cite{anderson84}. 
The Josephson effects (JE)
are a  paradigm 
of phase coherence in superfluid/superconductive systems 
\cite{anderson84,barone82}.
In the BEC context, JE have been associated with
coherent collective oscillations 
between two weakly-linked condensates, the weak link being created by
the potential barrier in a double-well trap 
\cite{sfgs-rsfs,zapata98,milburn97,dalfovo96,smerzi99,raghavan99,ruos-walls98}, 
or by a quasi-resonant external elettromagnetic field, which induces
oscillations between two trapping 
hyperfine levels \cite{williams99,villain99}.
Although bosons are neutral, and external circuits are obviously absent,
it is also possible to study the close analougs of the 
the ``ac'' and ``dc'' effects 
observed in superconducting
Josephson junctions \cite{giovanazzi99}.
Quite recently, an array of ``mesoscopic'' weakly 
linked condensates has
been created in \cite{anderson98}, each trap,  
located at the antinodes of
an optical standing wave,
containing $\sim 1000$ condensate atoms. 

The classical Josephson Hamiltonian (CJH) has been 
casted in term of a pendulum-like equation \cite{sfgs-rsfs,nota1}:
\begin{equation}
H_{cl} = {E_c \over 2} n^2 -
\frac{2 E_J}{N}  \sqrt{N^2/4 - n^2} \cos {\phi}
\label{eq:eq1}
\end{equation}
with the relative number of 
condensate atoms between the two bulk condensate
$n = \frac{1}{2} (n_1 - n_2)$ 
playing the role of the momentum,
and the relative phase $\phi = \phi_1 - \phi_2$ 
being the angle respect 
to the horizontal axis \cite{nota2}. 

The ``charging energy'' $E_c$ and the ``Josephson coupling energy''
$E_J$ can be calculated as overlap integrals \cite{sfgs-rsfs,giovanazzi99}:
\begin{mathletters}
\begin{eqnarray}
E_c&=&2 g \int d r ~\Phi _1( r)^4=
2 g \int~ dr~\Phi _2(r)^4  \\
E_J   &=& - N \int d r ~\Phi _1(r)
\left[ -\frac{\hbar ^2}{2m}{ \nabla}^2
+ V_{ext} +  \frac{g N}{2} [\Phi^2 _1(r) + \Phi^2 _2 (r)
] \right] \Phi _2(r)
\label{eq:eq2}
\end{eqnarray}
\end{mathletters}
with the one-body wave functions $\Phi_1(r), \Phi_2(r)$
localized in the trap $1,2$, respectively, and
$\int dr~\Phi _1(r)^\ast  \Phi _2(r) = 0$,
 $\int dr~|\Phi _{1,2}(r)|  = 1$, 
$g = \frac{4 \pi \hbar a}{m}$; $a$ being the scattering length and
$m$ the atomic mass; $N$ is the total number 
of atoms \cite{nota3}. 

The quantum Josephson Hamiltonian (QJH), on the other hand, 
has been studied within two different models 
which seem in conflict with each other
\cite{javanainen99}.
This conflict originates, at its heart, from a
fundamental, yet unsolved, problem: the existence 
and the meaning of a quantum phase operator \cite{susskind64}.

In the ``boson mode'' representation, QJH is written
in terms of creation and destruction boson operators
\cite{milburn97,raghavan99,ruos-walls98,villain99}:
\begin{equation}
\hat {H} = {E_c \over 4}  (\hat{a}^+_1 \hat{a}^+_1  \hat{a}_1 \hat{a}_1 +
\hat{a}^+_2 \hat{a}^+_2  \hat{a}_2 \hat{a}_2)
- \frac{E_J}{N} (\hat{a}^+_1 \hat{a}_2 + \hat{a}^+_2 \hat{a}_1) 
\label{eq:eq3}
\end{equation}
where $\hat{a}^+_{1,2}, (\hat{a}_{1,2})$ 
creates (destroys) a particle in the trap $1,2$, respectively.

In the ``phase'' representation, 
on the other hand,
the relevant quantum observables
are the difference of phases and number of atoms
between the two condensates in each trap 
\cite{smerzi99,leggett91,schon90}, the Hamiltonian 
being written in term of a ``quantum pendulum'' (Mathieu)
equation:
\begin{equation}
\hat{H} = -{E_c \over 2} \frac{\partial^2}{\partial \phi^2}
- E_J  \cos {\phi} 
\label{eq:eq4}
\end{equation}
with $\hat{n}^2 = ( \hat a_1^+ \hat a_1 -  \hat a_2^+ \hat a_2)^2 
 \sim -\frac{\partial^2}{\partial \phi^2} $
and $\hat \phi$ 
which cannot, as we will discuss shortly, be unambiguously defined in terms of
creation/destruction operators.
Eq.~(\ref{eq:eq4}) acts on a $2 \pi$-periodic 
``wave-function'' 
$\psi(\phi) = \psi(\phi+2 \pi)$, with 
${1 \over {2 \pi}} \int_{-\pi}^{\pi} 
d \phi  ~ \psi(\phi)^\ast~\psi(\phi) =1$ and
expectation values 
$\langle \hat A \rangle = 
{1 \over {2 \pi}} \int_{-\pi}^{\pi} 
d \phi  ~ \psi(\phi)^\ast~ A(\phi,t)~ \psi(\phi)$.
This framework,
generalized to include damping effects and external circuits,
is rather standard in the SJJ literature \cite{schon90}.

Generally speaking, the Hamiltonian 
Eq.~(\ref{eq:eq3}) can be derived microscopically from the
quantum field equation:
\begin{equation}
\hat H = \int dr~ \hat \Psi^+ \left[
-\frac{\hbar ^2}{2m}
\nabla^2 + V_{ext} + \frac{g}{2}~ \hat \Psi^+ \hat \Psi \right] \hat \Psi
\label{eq:eq5}
\end{equation}
writing the boson field in the ``two-mode approximation'' as: 
$\hat \Psi \simeq \Phi_1(r)~ \hat{a}_1 + \Phi_2(r)~ \hat{a}_2$. 

The quantum pendulum Eq.~(\ref{eq:eq4}), on the other hand, can be retrieved
quantizing the classical Josephson Hamiltonian 
Eq.~(\ref{eq:eq1}) by replacing the classically conjugate phase/atom number   
with the corresponding, non-commuting, operators \cite{nota4}.

On this grounds we can consider
Eq.~(\ref{eq:eq4}) as a phenomenological, while
Eq.~(\ref{eq:eq3}) has a clearer ``microscopical'' foundation:
it is generally believed that, for some limit, these two Josephson 
Hamiltonians 
describe essentially the same
physics, yet, their exact relation has not been exploited
in the literature.
A way to analyze, a posteriori, the relation 
between the two Hamiltonians Eq.~(\ref{eq:eq3}) and
Eq.~(\ref{eq:eq4}), is to identify:
\begin{equation}
\hat{a}_i \equiv e^{i \hat{\phi}_i} \hat{n}_i^{1/2}
\label{eq:eq6}
\end{equation}
with $\hat \phi_i$ and $\hat n_i = \hat a_i^+ \hat a_i$ 
Hermitian phase and number
operators acting in the trap $i$. 
The assumed Hermiticy of $\hat \phi_i$ implies the unitarity of 
$exp(i \hat{\phi_i})$.
Number and phase, thus, 
are treated as conjugate observables:
\begin{equation}
[\hat{\phi_i},\hat{n_i}] = i~~~~~~~~~~~~~~~~~~(?)
\label{eq:eq7}
\end{equation}
which implies, in the phase-representation, 
$\hat{n_i} \equiv - i \frac{\partial}{\partial \phi_i}, 
\hat \phi_i \equiv \phi_i$.  
This approach is known to be incorrect.  
The commutator gives rise to inconsistencies when its matrix elements 
are calculated in a number-state basis.
Even worse, the
exponential operator $exp(i \hat{\phi})$ is not unitary
and so does not define an Hermitian $\hat \phi$ \cite{susskind64}. 

Despite such problems, the Josephson Hamiltonian, Eq.~(\ref{eq:eq4}), 
is considered as 
the starting point of most analysis in SJJ and BJJ, the  
implicit caveat being that
the commutator relation Eq.~(\ref{eq:eq7}) is 
approximately correct for systems 
with a large number of ``condensate'' Cooper-pairs/atoms.
More precisely, the relation between Eq.~(\ref{eq:eq3}) 
and Eq.~(\ref{eq:eq4}) is considered only in a
semiclassical context, replacing
$\hat a_{1,2} \to n_{1,2} e^{i \hat{\phi_{1,2}}}$,
$n_{1,2}$ being $c$-numbers, in the term proportional
to the Josephson coupling energy $E_J$ in 
Eq.~(\ref{eq:eq3}).

In this letter we develops a consistent procedure to rewrite
the quantum Josephson Hamiltonian Eq.~(\ref{eq:eq3})
in the phase representation, overcoming the problem of defining
an Hermitian phase operator. The resulting equation
differs in several ways from the
Mathieu equation Eq.~(\ref{eq:eq4}), which is recovered in a specific limit.
Although we consider, for simplicity, the case
of a two-sites Josephson junction, our analysis
clarifies 
the relation between the Boson Hubbard \cite{jaksch98,amico99}
(which can be seen as the $n$-sites
generalization of Eq.~(\ref{eq:eq3}))
and the Quantum Phase models ($n$-sites
generalization of Eq.~(\ref{eq:eq4}))
\cite{otterlo95}, governing 
the dynamics of a BEC's array as that created in \cite{anderson98}.  

An arbitrary state spanning the two-dimensional Hilbert space,
in which is defined the QJH Eq.~(\ref{eq:eq3}),
can be expanded as:
\begin{equation}
\label{eq:eq8}
|\psi \rangle = {1 \over {(2 \pi)^2}} \int_{-\pi}^{\pi} 
d \phi_1 d \phi_2 ~ f(\phi_1, \phi_2) |\phi_1, \phi_2 \rangle
\end{equation}
with the (un-normalized) Bargmann state \cite{bargmann61}:
\begin{equation}
\label{eq:eq9}
|\phi_1, \phi_2 \rangle = \sum_{l,m=0}^{\infty} e^{i (l \phi_1 + m \phi_2)} 
{1 \over{\sqrt{l ! m !}}} |l \rangle |m \rangle
\end{equation}
with $|m \rangle, |l \rangle$ atom number eigenstates of trap $1,2$ 
respectively. 

In the Bargmann space, 
the action of the boson operators on the
state vector Eq.~(\ref{eq:eq8}) reduces to differential
operators acting on $f(\phi_1, \phi_2)$:
\begin{mathletters}
\label{eq:eq10}
\begin{eqnarray}
f(\phi_i)~ \hat a_i &\to& e^{i \phi_i}~ f(\phi_i) \\
f(\phi_i)~ \hat a^+_i &\to& i \frac{\partial}{\partial \phi_i}
 [e^{-i \phi_i} f(\phi_i)] 
\end{eqnarray}
\end{mathletters}
as can be seen after an integration by parts.
In particolar, the atom number operator assumes 
the familiar form: 
$f(\phi_i)~\hat n_i = f(\phi_i)~ \hat a^+_i \hat a_i \to 
- i {\partial \over {\partial \phi_i}}~ f(\phi_i)$.

We consider as fixed the total number of atoms, and we can write: 
\begin{equation}
f(\phi, \phi_+) = e^{-i N \phi_+} \psi(\phi)
\label{eq:eq11}
\end{equation}
with $\phi_+ = {1 \over 2} (\phi_1 + \phi_2),
~\phi = \phi_1 - \phi_2$,
$N = n_1 + n_2, ~n = {1 \over 2} (n_1 - n_2)$, $N$ and $n$ being the 
total and the relative number of atoms. 
$\psi(\phi)$ is an arbitrary, $2 \pi$-periodic function 
normalized to unity: $ \frac{1}{2 \pi}
\int_{-\pi}^{\pi} d \phi   |\psi(\phi)|^2 = 1$. We can
integrate over $\phi_+$ the Eq.~(\ref{eq:eq8}):
\begin{equation}
|\psi \rangle = {1 \over {2 \pi}} \int_{-\pi}^{\pi} 
d \phi  ~ \psi(\phi) |\phi \rangle 
\label{eq:eq12}
\end{equation}
with 
\begin{equation}
|\phi \rangle = \sum_{n= -N/2}^{N/2}  
{{ exp(i n \phi)} \over{\sqrt{({N \over 2} + n)! {({N \over 2} - n) !}}}} 
|n \rangle
= N ! 
(\hat a_1^+ e^{i {\phi \over 2}} + a_2^+ e^{-i {\phi \over 2}})^N |0 \rangle
\label{eq:eq13}
\end{equation}
that can be considered, in the large $N$ limit, as a pure phase state
\cite{leggett91}.
The state Eq.~(\ref{eq:eq13}) is proportional (with $\phi = 0$)
to the exact ground state of Eq.~(\ref{eq:eq3}) in the non-interacting
limit ($E_c = 0$).
Here we consider the
states $|\phi \rangle$ as an overcomplete base to expand
the solution of the full Josephson Hamiltonian Eq.~(\ref{eq:eq3}).
The scalar product of two phase states is given by: 
\begin{mathletters}
\begin{eqnarray}
\langle \psi|\psi \rangle &=&  {1 \over {(2 \pi)^2}} \int_{-\pi}^{\pi} 
d \theta~ d \phi~ \psi(\theta)^\ast~ \psi(\phi)~  
\langle \theta|\phi \rangle =1\\ 
\langle \theta|\phi \rangle &=& 
\sum_{n= -N/2}^{N/2} 
{e^{i n (\phi-\theta)}\over{({N \over 2} + n)! {({N \over 2} - n) !}}} = 
\frac{2^N}{N!} {\cos}{^N}({{\phi - \theta}\over 2}) = 
2 \pi \frac{2^N}{(N !!)^2}~ \delta_N(\phi - \theta) 
\label{eq:eq14}
\end{eqnarray}
\end{mathletters}
In the limit of large number of atoms, 
$\langle \theta|\phi \rangle$ becomes proportional (in the interval 
$-\pi \leq \phi, \theta < \pi$) to a delta function: 
\begin{equation}
\lim_{N \to \infty} \delta_N(\phi - \theta) =
\delta(\phi - \theta)
\label{eq:eq15}
\end{equation}
The action of the Josephson Hamiltonian Eq.~(\ref{eq:eq3}) on the state vector
Eq.~(\ref{eq:eq12}) gives:
\begin{mathletters}
\label{eq:eq16}
\begin{eqnarray}
\hat{H} |\psi \rangle &=& (\hat H_c + \hat H_J) |\psi \rangle \\
\hat{H}_c |\psi \rangle &=& {E_c \over 4}  {1 \over {2 \pi}} 
\int_{-\pi}^{\pi} d \phi~ |\phi \rangle~
[ - 2 \frac{\partial^2}{\partial \phi^2} + 
( {1 \over 2} N^2 - N) ] ~ \psi(\phi) \\
\hat{H}_J |\psi \rangle &=& - \frac{E_J}{N} {1 \over {2 \pi}} 
 \int_{-\pi}^{\pi} d \phi~  |\phi \rangle ~
[ (N + 2) \cos \phi + 
2 \sin \phi~ \frac{\partial}{\partial \phi} ] 
~\psi(\phi)
\end{eqnarray}
\end{mathletters}
Projecting out in the phase space:
\begin{mathletters}
\label{eq:eq17}
\begin{eqnarray}
\langle \theta| \hat{ H}|\psi \rangle &=& 
\frac{1}{2 \pi} \int_{-\pi}^{\pi}~ 
d \phi~ \hat{F}~ \psi(\phi) ~ \langle \theta|\phi \rangle\\ 
\hat{F}~ \psi(\phi)  &\equiv& [- \frac{E_c}{2} 
\frac{\partial^2}{\partial \phi^2} 
- E_J (1 + \frac{2}{N}) \cos\phi~  -
 2 \frac{E_J}{N}~ \sin{\phi}~ \frac{\partial}{\partial \phi}]~ \psi(\phi)
\end{eqnarray} 
\end{mathletters}
where a constant energy shift has been omitted.
The spectrum of $\hat{H}$ is given by the eigenvalue equation for this 
$\hat{F}$, which is known as the 
Ince equation \cite{arscott64}.
Notice that, because of the overcompleteness of the phase-states 
Eq.~(\ref{eq:eq13}),
$\hat H_J$ in Eq.~(\ref{eq:eq16},c) is Hermitian,  
even though the differential operator 
$\sin{\phi}~ \frac{\partial}{\partial \phi}$ is not. 

Thus the time-dependent Schr\"odinger equation in our phase representation, 
\begin{equation} 
i \frac{\partial }{\partial t} \langle \phi|\psi \rangle 
- \langle \phi|(\hat{H_c} + \hat H_J) |\psi \rangle =  0
\;,
\label{eq:eq18}
\end{equation}
can be put in dimensionless form (by defining 
$\gamma = \frac{2 E_J}{E_c} \frac{1}{N}$ and rescaling 
$t \to t\frac{E_c}{2}$) as
\begin{equation}
\int_{-\pi}^{\pi} d \phi~\delta_N(\phi - \theta)~  
[ i \frac{\partial}{\partial t}
+ \frac{\partial^2 }{\partial \phi^2} 
+ \gamma (N + 2) \cos\phi~  +
 2 \gamma~ \sin{\phi}~ \frac{\partial}{\partial \phi}]
~ \psi(\phi) = 0
\label{eq:eq19}
\end{equation} 
We may obtain a more transparent form of (\ref{eq:eq19}) by
defining $\psi(\phi) = e^{\gamma \cos \phi} \Psi (\phi)$.
This eliminates the first derivative and, dropping a constant, maps $\hat{F}$ 
onto a manifestly Hermitian operator acting on $\Psi$.  The result is that
to satisfy Eq.~(\ref{eq:eq19}) it is sufficient to satisfy
\begin{equation}
i \frac{\partial \Psi}{\partial t} = 
[- \frac{\partial^2}{\partial \phi^2} 
- \gamma (N + 1) \cos{\phi} -
\frac{1}{2} \gamma^2 \cos{2 \phi} ] \Psi
\;.
\label{eq:eq21}
\end{equation}   
Setting $i\frac{\partial \Psi}{\partial t} = E\Psi$ gives the three-term Hill 
equation \cite{arscott64}.

Eq.~(\ref{eq:eq21}) is of the same usefully simple form as the QPM.  
One can apply to it all the intuition and experience,
and all the calculational tools, that are relevant to the one-particle
Schr\"odinger equation with a periodic potential.  Solving it yields
the energy spectrum of the BHM, and thus any dynamical time scales (such as for
decay of the possible metastable states by tunneling).  But since
the equation is based on an overcomplete representation, expectation values
and matrix elements must be computed using the nonstandard inner product with
the kernel $\delta_N$, and with the extra $e^{\gamma\cos\phi}$ factors; hence
the interpretation of the eigenstates can be somewhat subtle.  
%
%
Since $\delta_N$, with $N>>1$, behaves
as a delta function for test functions varying slowly on the scale $N^{-1/2}$,
for the lower energy states the inner product is essentially the standard
one, and Eq.~(\ref{eq:eq21}) can be used without any extra thought.  For the
higher energy states, however, the interpretations of $\Psi(\phi)$ and 
Eq.~(\ref{eq:eq21}) depend on $\gamma$.

We can distinguish three regimes.  For $\gamma\geq {\cal O}(N)$, the $\cos2\phi$
term in (\ref{eq:eq21}) is a significant correction to the purely $\cos\phi$ 
potential
of the QPM; and for $\gamma > N/2$ it makes a second local minimum at 
$\phi=\pi$.
This agrees with the GPE result that for $\gamma>N/2$ there are metastable 
$\pi$-
oscillations of zero average $n$.  But the trapping of probability near 
$\phi=\pi$
is not trivial, because $e^{\gamma\cos\phi}$ factor tends strongly to suppress 
it.
This suppression is countered for high energy states, however, by the fact that
$\delta_N$ effectively eliminates all Fourier components $e^{ik\phi}$ with 
$|k|>N/2$ 
in $\psi(\phi)$.  (This is as it should be, since from our derivation it is 
clear 
that such components are unphysical.)  There are thus high energy solutions to 
Eq.(\ref{eq:eq21}) whose physical components are localized around $\phi=\pi$, 
because
in the deeper well around $\phi=0$ their WKB frequency has $|k|>N/2$.  In the 
intermediate range of energies, however, the implications of $\delta_N$
are more complicated.  Eqn.~(\ref{eq:eq21}) may still be treated with 
semiclassical
methods, as will be discussed elsewhere.  It is qualitatively clear that both
the second minimum of the potential, and the failure of quantum motion to be
strictly confined to it, will be important; and so we must consider 
(\ref{eq:eq21})
as giving corrections to both the QPM and the GPE.  

For $1<\gamma<<N$, the $\cos2\phi$ term is insignificant, and so the 
eigenfunctions
of Eq.~(\ref{eq:eq21}) are essentially the Mathieu functions of the QPM.  The 
nonlocality
of $\delta_N$ is of little qualitative significance, except for the highest 
energy
states.  These are well approximated by the WKB method, and their local wave
number is of the form $k + \gamma{N+1\over2k}\cos\phi$.  For $k$ of order $N/2$,
this implies that solutions to (\ref{eq:eq21}) of the `running' type, with
no localized phase, actually have their physical component localized around 
$\phi=\pi$.  This supports the GPE in what is its only significant discrepancy 
with 
the QPM in this range of $\gamma$, for the GPE also predicts high energy 
$\pi$-states 
with nonzero average $n$\cite{sfgs-rsfs}.  (We neglect the fact that the actual 
quantum 
eigenstates
must have zero average $n$, because they are even and odd `Schr\"odinger's Cat' 
superpositions of the states with large positive and negative $n$.  But the 
splitting
is exponentially small in $N$.)

For $\gamma\leq 1$, the situation is similar, except that the modulation
of the local wave number with $\phi$ is too small for the elimination of 
unphysical 
Fourier components to produce a localized state.  It simply eliminates all WKB
states with $|k|>N/2$, and the result is that the physical spectrum of 
Eq.~(\ref{eq:eq21}) is essentially Mathieu functions, up to a cut-off.  The 
$\pi$-states
of the GPE are absent and the QPM is vindicated, except
for the welcome truncation of the spectrum, in accordance with the fact that
the Hilbert space must indeed be only $(N+1)$-dimensional.

In  Fig.~(\ref{fig1}) we draw
the probability in phase, $P(\phi)$, in an orthonormal representation of
$\phi$. This illustrate the interpretation of our nonorthonormal phase
representation with unphysical components projected out.
Considering $N=199$
particles and $\gamma=N/5$, all 200 energy eigenstates are found
numerically in the number basis, and ranked in order of energy.  The
absolute square of the Fourier transform of the number amplitudes is
plotted for each state; since all $P(\phi)$ are even, only
$0\leq\phi\leq\pi$ is shown.
This approach is not possible for large $N$, but it illustrates and
confirms our discussion of the physical interpretation of our
overcomplete representation, which is useful for all $N$.  For this
value of $\gamma$, Josephson-localized, running, and $\pi$-localized
states all occur.

To conclude, a few words are in order about the Rabi limit $\gamma\to\infty$.
In this case  the potential of Eq. (20) becomes a
pair of strong harmonic potentials centred at $\phi=0$ and $\phi=\pi$,
both having the frequency $\gamma E_c = 2 E_J/N$. The equally spaced
harmonic oscillator eigenstates of these potentials, up to the cut-off
described earlier, reproduce the Rabi spectrum of the Hamiltonian
(3) in the limit $E_c \to 0$.

This work has been partially supported by the Cofinanziamento MURST.

\begin{figure}
\caption{
The module square of the energy eigenstates $P(\phi)$, 
as a function of an orthonormal 
representation of $\phi$, with $N=199$ and $\gamma=N/5$. 
}
\label{fig1}
\end{figure}


\begin{thebibliography}{10}
\bibitem{nota1} Here and in the following we consider a BEC in a symmetric
double-well potential. In the case of a asymmetric potential we should
add a term proportional to $n$
that is irrelevant for our present porpoises.
Terms proportional to  $\int d r ~\Phi _1( r)^2~
\Phi _2(r)^2$ are also neglected.
\bibitem{nota2}
Peculiariar is the
``non-rigidity'' of the pendulum, which has a momentum 
dependent length $\sim  \sqrt{N^2/4 - n^2} $. This allows stable
pendulum oscillations about the up-right position, i.e. $\pi$-oscillations.
\bibitem{nota3} It is clear, in this context, that the Josephson 
Hamiltonian Eq.~(\ref{eq:eq1}) is ``classical'' in the sense that
it considers the conjugate relative phase $\phi$ and number of atoms $n$ 
as c-numbers.
\bibitem{nota4} However this formal replacement can be pursued 
only in the limit $n << N; E_c >> E_J/N^2$ so to keep only the zero-order 
expansion term of the square-root in Eq.~(\ref{eq:eq1}).
\end{thebibliography}
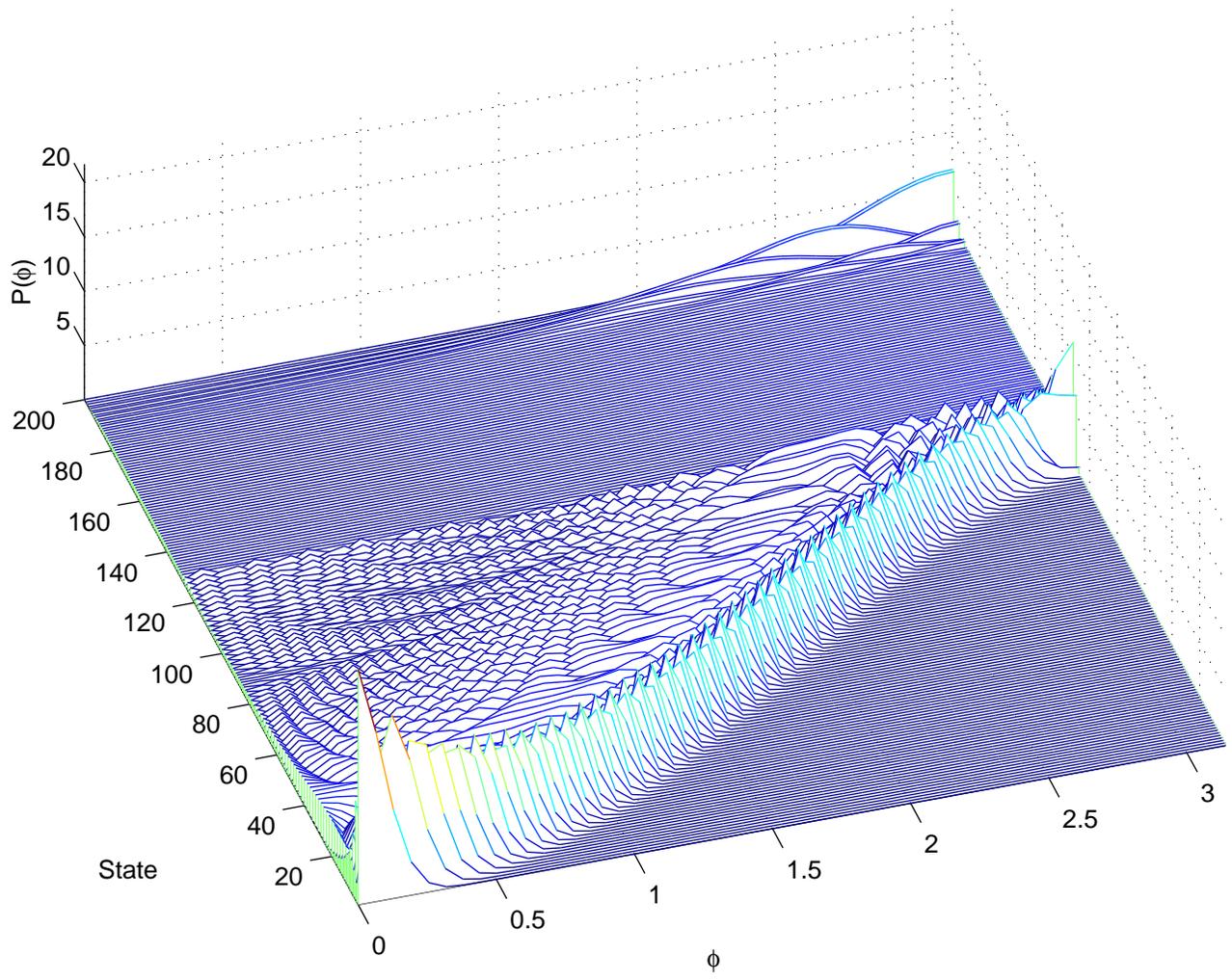
\end{document}